\documentclass[fleqn,usenatbib]{mnras}

\usepackage{newtxtext,newtxmath}

\usepackage[T1]{fontenc}
\usepackage{ae,aecompl}
\usepackage{xcolor}

\usepackage{graphicx}	
\usepackage{amsmath}	

\def\msun{\rm{M}_{\odot}}
\def\arcsec{^{\prime\prime}}

\title[DM cross-section from radial arcs]{Constraining the cross-section of dark matter with giant radial arcs in galaxy clusters}

\author[J. Vega-Ferrero et al.]{J. Vega-Ferrero,$^{1,2}$\thanks{E-mail: vegaj@ifca.unican.es} 
J. M. Dana,$^3$
J. M. Diego,$^1$
G. Yepes,$^{4,5}$
W. Cui$^{6}$ and
\newauthor
M. Meneghetti$^{7,8}$\\\\
$^1$IFCA, Instituto de F\'{\i}sica de Cantabria (UC-CSIC), Av. de Los Castros s/n, 39005 Santander, Spain \\
$^2$Department of Physics and Astronomy, University of Pennsylvania, 209 S. 33rd St, Philadelphia, PA 19104, USA\\
$^3$ Almer\'{\i}a, Spain\\
$^4$Departamento de F\'{\i}sica Te\'orica M-8, Universidad Aut\'onoma de Madrid, Cantoblanco, E-28049 Madrid, Spain\\
$^5$ Centro de Investigaci\'on Avanzada en F\'{\i}sica Fundamental (CIAFF), Universidad Aut\'onoma de Madrid, 28049 Madrid, Spain\\
$^6$ Institute for Astronomy, University of Edinburgh, Royal Observatory, Edinburgh EH9 3HJ, United Kingdom \\
$^7$INAF - Osservatorio di Astrofisica e Scienza dello Spazio, via Gobetti 93/3, 40129, Bologna, Italy \\
$^8$INFN, Sezione di Bologna, viale Berti Pichat 6/2, 40127, Bologna, Italy}

\date{Accepted 2020 October 14. Received 2020 September 18; in original form 2020 June 5}
\pubyear{2020}

\begin{document}
\label{firstpage}
\pagerange{\pageref{firstpage}--\pageref{lastpage}}
\maketitle

\newcommand\ls{\ensuremath{\hbox{\rlap{\hbox{\lower4pt\hbox{$\sim$}}}\hbox{$<$}}}}
\newcommand\gs{\ensuremath{\hbox{\rlap{\hbox{\lower4pt\hbox{$\sim$}}}\hbox{$>$}}}}

        
\begin{abstract}
We compare the statistics and morphology of giant arcs in galaxy clusters using N-body and non-radiative SPH simulations within the standard cold dark matter model and simulations where dark matter has a non-negligible probability of interaction (parametrized by its cross-section), i.e self-interacting dark matter (SIDM). We use a ray-tracing technique to produce a statistically large number of arcs around six simulated galaxy clusters at different redshifts. Since dark matter is more likely to interact in colliding clusters than in relaxed clusters, and this probability of interaction is largest in denser regions, we focus our analysis on radial arcs (which trace the lensing potential in the central region better than tangential arcs) in galaxy clusters which underwent (or are undergoing) a major merger. We find that self-interacting dark matter produces fewer radial arcs than standard cold dark matter but they are on average more magnified. We also appreciate differences in the arc morphology that could be used to statistically favor one model versus the other. 
\end{abstract}

\begin{keywords}
gravitational lensing; galaxies: clusters: general; cosmology: theory - dark matter
\end{keywords}


\section{Introduction}\label{sec:intro}

The nature of dark matter is arguably one of the biggest mysteries of modern science. Despite the wealth of observational evidence for its existence (from astrophysical probes), all efforts for its direct detection have proven fruitless. The lack of success in detecting the elusive dark matter has promoted the appearance of alternative explanations, including those that propose that dark matter does not exist, and all astrophysical evidences can be reinterpreted with a reformulation of the laws gravity. However, no model has been as successful at reproducing cosmological observations as the particle dark matter (DM) model where DM behaves like collision-less cold DM at large cosmological scales, the so-called Cold Dark Matter (CDM). This includes the cosmic microwave background power spectrum, which can be reproduced with astonishing precision only when a precise amount of CDM is included in the model. 

In addition to the efforts being made for direct detection (or production) of dark matter, it is important to pursue the detection of dark matter through indirect methods provided by the astrophysical probes like decay or annihilation into gamma-rays, other particles or EM radiation, accumulation effects on astrophysical bodies (such as stars or neutron stars), distortions in the cosmic microwave background, gravitational lensing, etc. \citep[see][for a review]{Gaskins2016}.

In this context, the CDM cosmological model has been extremely successful in describing the large scale structure that we observe of our Universe. However, at smaller scales, where the formation of structures becomes non-linear, the CDM scenario has found difficulties explaining some discrepancies (such as the core-cusp, the diversity, the missing satellites and the too-big-to fail problems) that arose from a comparison with predictions from N-body cosmological simulations \citep{Bullock2017}. A promising alternative to the collision-less CDM cosmological model is the self-interacting dark matter (SIDM) model, proposed by \citep{Spergel2000} to solve both the core-cusp and the missing satellites problems. In this framework, DM particles scatter elastically with each other through $2 \rightarrow 2$ interactions. Since the scattering rate of DM particles is proportional to the DM density and their relative velocities, the SIDM model remains successful on large scales (almost identical to the CDM framework), but changing the formation of structures at late times and only on small scales, in particular, in the inner regions of DM halos. For instance, the relatively shallow profiles in the central regions of some galaxies and clusters (core-cusp problem) or the too-big-to fail problem could be all explained if the cross-section of DM per unit mass is around $\sigma/\rm{m} \approx 1~\rm{cm}^2/\rm{g}$ \citep[see][for an exhaustive review]{Tulin2018}.

Among the different astrophysical probes, galaxy clusters have provided useful information about dark matter properties. In the proposed SIDM cosmological model (with a velocity-dependent scattering rate), the self-interaction rate for clusters is expected to be much larger than for galaxy scales, given that the DM densities and the typical scatter velocity (of the order of $v_{\rm{rel}} \approx 1000~\rm{km/s}$) are larger in massive DM halos. N-body cosmological simulations have been an indispensable tool to study the effects of DM interactions in structure formation on a wide range of scales, from dwarf galaxies to massive galaxy clusters. Recently, N-body simulations including SIDM with high resolution and halo statistics have reactivate the SIDM scenario as a plausible alternative to the CDM cosmological model \citep{Vogelsberger2012,Rocha2013,Peter2013,Zavala2013,Vogelsberger2014,Elbert2015,Fry2015,Dooley2016,Wittman2018,Robertson2019}. Although more recent studies based on massive clusters point towards smaller values of the cross-section of DM per unit mass ($\sigma/m \lesssim 0.1~\rm{cm}^2/\rm{g}$) on these scales \citep{Kaplinghat2014,Elbert2015}, other studies based on stacked merging clusters \citep{Harvey2015} and stellar kinematics within cluster cores \citep{Elbert2015} suggest some tension with these values. Additionally, as shown in \citet{Randall2008}, the Bullet Cluster shows how dark matter is consistent with the hypothesis that it is collision-less. Along with other merging clusters, the Bullet Cluster has provided an upper limit for the cross-section of DM per unit mass at $\sigma/m \approx 1~\rm{cm}^2/\rm{g}$. Larger values of the cross-section would result in a shift between the peak of the dark matter distribution and the centre of the distribution of galaxies (galaxies behave like truly collision-less particles even during a major cluster merger). Such a shift was not observed by \citet{Clowe2006}, suggesting that the cross-section of DM per unit mass must be $\sigma/m \approx 1~\rm{cm}^2/\rm{g}$ at most. However, \citet{Robertson2017} found substantially larger DM--galaxy offsets that suggest even a larger value of $\sigma/m = 2~\rm{cm}^2/\rm{g}$. Besides, Abell 3827 and Abell 520 exhibit large DM-stellar offsets that, if explained by the SIDM model, will require larger DM cross-sections probably in tension with other data, such as the core sizes in galaxy clusters.

Alternatively, strong gravitational lensing data has been used to constrain the core size and density in clusters relevant for SIDM \citep{Firmani2000,Firmani2001,Meneghetti2001,Wyithe2001}. More recent studies have used Einstein radii statistics in galaxy clusters to constrain the DM cross-section \citep{Robertson2019}, suggesting that future wide surveys might be able to distinguish between CDM and SIDM cosmological models \citep{Despali2019}.

In particular, \citet{Meneghetti2001} placed the strongest constraint on the DM cross-section and cluster cores by examining the ability of a SIDM halo to produce "extreme" strong lensing arcs (both radial or giant tangential arcs). The authors concluded that their SIDM halo simulated with $\sigma/m > 1~\rm{cm}^2/\rm{g}$ is not dense enough (in projection) to produce extreme tangential arcs with length-to-width ratios of $l/w \gtrsim 3.5$. Based on the ability to produce radial arcs, the constraint found is even more severe, since their SIDM halo simulated with $\sigma/m < 0.1~\rm{cm}^2/\rm{g}$ was not able to produce radial arcs. As acknowledged in \citet{Meneghetti2001}, constraints based on one single halo need to be addressed with caution due to the variability of the density profiles in SIDM halos. Moreover, this particular halo is a SIDM-only simulation (no baryons were included) and, therefore, it did not account for the baryonic density due to the central galaxy which boosts its lensing efficiency. One of the major effects of the cooling and the star formation in simulations is to trigger a strong adiabatic contraction of the baryonic component, leading to (sometimes unrealistically) denser cluster cores. Nevertheless, there have been found several examples of galaxy clusters that exhibit radial arcs. For instance, 12 candidate radial arcs where found in three of the six clusters examined by \citep{Sand2004,Sand2005}. More recent studies by \citet{Newman2013a,Newman2013b} have found radial arcs in two of the clusters analyzed by \citet{Sand2004}, MS2137-23 and Abell 383, with DM cores (of the order of 10 kpc) which are consistent with lensing data once the baryonic mass is included. Another interesting example with several central images detected (and $\sim30\%$ of its 82 multiple images classified as radial arcs) is MACS J1206 \citep{Caminha2017}. Interestingly, as first proposed by \citet{Molikawa2001}, it is practical to use the ratio of radial to tangential arcs as a statistical measure of the slope of the dark matter distribution in cluster cores \citep[see also][]{Sand2005}.

In this context, it is important to determine to what extend the constraint of $\sigma/m < 0.1~\rm{cm}^2/\rm{g}$ found in \citet{Meneghetti2001} is still valid. We address this issue by producing a new set of N-body/SPH simulations of several cluster-scale DM halos, at different redshifts, and including the presence of baryons in both CDM and SIDM frameworks. We study the capacity of these halos to produce elongated radial arcs (among other properties). By comparing our results with observations, one can set limits on $\sigma/m$, by exploiting the properties of radial arcs (such as their lengths, widths, convergences, shears and magnifications).

This work is organized as follows. Section~\ref{sec:sims} presents the set of N-body/SPH cosmological simulations of cluster-size DM halos for both CDM and SIDM cosmological models and a summary of overall properties, such as masses and shapes. Section~\ref{sec:lensing} describes the gravitational lensing properties of the simulated clusters by means of Einstein radii statistics and the formation of radial arcs. Finally, section~\ref{sec:discussion} summarizes our main results and conclusions.

\section{Numerical simulations}
\label{sec:sims}
In this study, we analyze a total of six massive galaxy clusters extracted from two sets of well tested, high-resolution N-body/SPH cosmological simulations with different cosmological parameters. For each cluster we compare the results obtained with the simulation particles set as standard CDM particles and an identical simulation (i.e, with the same initial condition) where the DM particles are allowed to interact with a given probability determined by the value of $\sigma/m < 1~\rm{cm}^2/\rm{g}$ (SIDM). 

\subsection{MUSIC-MD simulations}
\label{sec:MUSIC}

One of the galaxy clusters used in this study is extracted from the MUltidark SImulations of galaxy Clusters (MUSIC\footnote{\url{http://music.ft.uam.es}}, \citealt{Sembolini2013}). In particular, we analyze the MUSIC-MD dataset, which consists of a set of re-simulated clusters extracted from the MultiDark Simulation\footnote{\url{https://www.cosmosim.org}} (MDR1, \citealt{Prada2012}), a DM-only simulation with $2048^3$ particles in a cubic box of $1h^{-1}$Gpc side. The MUSIC-MD simulation was done using the best-fit cosmological parameters to WMAP7 + BAO + SNI (\citealt{Komatsu2011}, $\Omega_M=0.27$, $\Omega_b=0.0469$, $\Omega_{\Lambda}=0.73$, $\sigma_8=0.82$, $n_s=0.95$, $h=0.7$). The MUSIC-MD clusters were selected according to a mass limited selection, taking all clusters within the MDR1 simulation with masses above $10^{15}h^{-1}\rm{M}_{\odot}$ at $z=0$. In total, 283 different Lagrangian regions corresponding to spheres of $6h^{-1}$Mpc radius were re-simulated with $4096^3$ particles centered on the most massive clusters found in the MDR1 simulation. Therefore, the mass resolution for the re-simulated clusters is 8 times larger than in the parent MDR1 simulation, that is $m_{\rm{DM}} = 9.01 \times10^8 h^{-1} \msun$ for the DM particles and $m_{\rm{SPH}} = 1.9 \times 10^8 h^{-1} \msun$ for the gas particles.

The MUSIC-MD clusters have been performed using the parallel \textsc{gadget2} Tree-PM code \citep{Springel2005} with both radiative and non-radiative hydrodynamics formulation for the gas particles. By comparing simulations with different treatments of baryonic processes, \citet{Killedar2012} found that the inclusion of gas cooling, star formation and AGN feedback together lead to lensing cross-sections that are similar to those obtained from simulations including only DM and non-radiative gas. For this reason and in order to avoid any artificial lensing boost due to the treatment of the baryonic processes, in this study, we only examine the non-radiative run of the MUSIC-MD simulations.

\subsection{\textit{The Three Hundred} project}
\label{sec:300}

As a second set of simulated galaxy clusters, we also analyzed 324 spherical regions centered on each of the most massive clusters ($\rm{M}>10^{15}h^{-1}\rm{M}_{\odot}$) identified at $z=0$ within the DM-only MultiDark simulation (MDPL2, \citealt{Klypin2016}). The MDPL2 simulation was performed using the cosmological parameters presented by \citet{Planck2016} ($\Omega_M=0.307$, $\Omega_b=0.048$, $\Omega_{\Lambda}=0.693$, $\sigma_8=0.823$, $n_s=0.96$, $h=0.678$). The MDPL2 is a periodic cube with a comoving length of $1h^{-1}$Gpc containing $3840^3$ DM particles. DM particles within the highest resolution Lagrangian regions are split into DM and gas particles, according to the assumed cosmological baryon fraction. The re-simulated clusters have a mass resolution of $m_{\rm{DM}} = 1.27 \times10^9 h^{-1} \msun$ for the DM particles and $m_{\rm{SPH}} = 2.36 \times 10^8 h^{-1} \msun$ for the gas particles. The radius of the spherical regions where the re-simulated clusters are centered is $15h^{-1}$Mpc and, therefore, much larger than their virial radius. The 324 galaxy clusters within \textit{The Three Hundred}\footnote{\url{http://the300-project.org}} project \citep{Cui2018,300Wang2018,300Mostoghiu2019,300Arthur2019,300Haggar2020,300Ansarifard2020,300Li2020,300Knebe2020,300Kuchner2020} were re-simulated using two different codes smooth-particle hydrodynamics (SPH) to follow the evolution of the gas component: the \textsc{gadget-music} code \citep{Sembolini2013} and 'modern' SPH code \textsc{gadget-x} \citep{Murante2010,Rasia2015}. Both codes are based on the gravity solver of the \textsc{gadget3} Tree-PM code (an updated version of the \textsc{gadget2} code; \citealt{Springel2005}), but they apply different SPH techniques as well as rather distinct models for the sub-resolution physics \citep[see][for more details]{Cui2018}. On one hand, the \textsc{gadget-music} run is performed using non-radiative SPH formulation and, therefore, gas particles can be heated only via gravitational collapse. While, on the other hand, the \textsc{gadget-x} includes an improved SPH scheme with AGN feedback and black hole seeding and growth. In this study, we only examine the non-radiative run (\textsc{gadget-music}) of the clusters within the \textit{The Three Hundred} project.

\subsection{Simulations of self-interacting DM halos}
\label{sec:sidm}

The aim of this study is to examine the effects caused by DM self-interactions on cluster-size DM halos extracted from cosmological numerical simulations. More specifically, we aim at a fiducial comparison of several morphological, dynamic and gravitational lensing features of cluster-scale DM halos.

Using a ray-shooting pipeline \citep[see][and references therein]{Meneghetti2010} we derive the gravitational lensing properties (such as deflection angles, convergence, shear and magnification maps) of all the DM halos with M$_{vir} > 2 \times 10^{14} h^{-1}\rm{M}_{\odot}$ within the MUSIC-MD dataset and of the most massive DM halos in the 324 re-simulated regions within the \textsc{gadget-music} run of \textit{The Three Hundred} project. For our lensing analysis of the MUSIC-MD dataset, we analyze the simulation snapshots at redshift $z = (0.250, 0.333, 0.429, 0.667)$ for 500 random projections along the line of sight \citep[see][for a detailed description]{Meneghetti2014}. We select the cluster 11 (hereafter clus11) for being the cluster with the largest effective Einstein radius ($\theta_E$) within the MUSIC-MD dataset, which is a good estimate of the lensing efficiency of a cluster lens. This selection is motivated not only by the fact that clus11 has a strong lens at $z=0.333$, but also by the fact that galaxy clusters at redshifts $0.2 \lesssim z \lesssim 0.4$ are the most efficient gravitational lenses for sources at redshifts $z_s \gtrsim 1$. Clus11 in the MUSIC-MD simulations shows an Einstein radius with $\theta_E \simeq 56$ arcsec for a cluster redshift of $z_l = 0.333$ and a source redshift of $z_s = 2.0$.

For the lensing analysis performed over the \textsc{gadget-music} dataset, we use the same ray-shooting pipeline as for the MUSIC-MD dataset to analyze four different projections along the line of sight (three of them are arbitrary, $x$-axis, $y$-axis and $z$-axis, and one corresponds to the cluster's major axis projected along the line of sight) for each of the 324 clusters at redshifts $z = (0.250, 0.333, 0.429, 0.538)$. Generally, we expect a larger strong lensing signal when the cluster mass distribution is projected along its major axis. Following the previous procedure, we select the clusters with the largest Einstein radii at each given redshift. Some of them are selected not only at one redshift and, therefore, we end up with a total of five galaxy clusters extracted form the \textsc{gadget-music} dataset, which are labelled as clus2, clus7, clus9, clus30 and clus82 . 

Then, we re-simulate clus11 and clus2, clus7, clus9, clus30 and clus82 for both CDM and SIDM cosmological models using the N-body and non-radiative SPH framework GIZMO \citep{Hopkins2015} with the same initial conditions and cosmological parameters as for the MUSIC-MD and \textit{The Three Hundred} simulations, respectively. It is important to note that we do not expect significant differences in the gravitational lensing properties of the cluster-size halos here presented due to the differences in the cosmological parameters between the clus11 extracted from the MUSIC-MD dataset and the other five halos extracted from the \textit{The Three Hundred} simulations. We also checked that the results obtained with GIZMO for the six mentioned cluster-size halos within the CDM model are consistent with the original simulations performed with the \textsc{gadget} code. In the SIDM model, DM particles scatter elastically with each other with a velocity independent cross-section of $\sigma / m = 1$ cm$^2/$g. Clus11 is a cluster-size DM halo that is undergoing a major merger between $z=0.300$ and $z=0.333$, with two DM clumps separated less than 500$h^{-1}$kpc at those redshifts. This situation is of particular interest for a detailed study of the DM self interactions, given the high rate of interactions that are expected to take place in the collision or merger of massive DM halos. Therefore, we simulate clus11 at three different redshifts, $z=(0.250,0.300,0.333)$, for both cosmological models. Moreover, clus2 is a strong lens at the four redshifts analyzed, $z = (0.250, 0.333, 0.429, 0.538)$; clus7 is one of the most efficient lenses at $z = 0.429$; clus9 is one of most efficient lenses at $z = (0.250, 0.333, 0.538)$; clus30 is one of the most efficient lenses at $z = 0.538$; and finally, clus82 is a strong lens at $z= 0.429$.

\subsubsection{Masses and shapes of DM halos}
\label{sec:shapes}

To characterize the halo in terms of its mass and shape we compute its triaxial shape following a similar procedure as the one described in \citet{Despali2013}. First, we assume a spherical overdensity (SO) criterion to find the virial radius and mass enclosing an average overdensity $\Delta_c \rho_c (z)$, with $\Delta_c$ defined as a certain overdensity value and $\rho_c (z)$ being the critical density of the Universe at a given redshift ($z$). Then, for particles found within the SO virial radius, we derive the mass tensor $M_{\alpha \beta}$ as follows:

\begin{equation}
M_{\alpha \beta} = \frac{1}{M_{vir}} \sum^{N_{vir}}_{i=1} m_i \mathbf{r}_{i,\alpha} \mathbf{r}_{i,\beta},
\label{eq:mtensor}
\end{equation}
where $M_{vir}$ is the SO virial mass, $N_{vir}$ is the number of particles within the SO virial radius, $m_i$ is the particle mass, $\mathbf{r}_i$ is the position vector of the $i$th particle and $\alpha$ and $\beta$ are the tensor indices ($x$, $y$ and $z$ components of the three coordinate axes). The mass tensor defined in this way allows to determine the halo shape for different types of particles (i.e., with different masses). Therefore, it is possible to compute the halo shape for the DM and GAS distribution separately, or the overall halo shape including both particles types.

\begin{table*}
\begin{center}
\caption{Summary of simulated halo properties and redshifts. First column corresponds to the cluster label. Second column denotes the cosmological model used for the simulation, either CDM or SIDM. Third column shows the redshift at which the different clusters have been analyzed. Fourth and fifth columns show the masses (in $10^{15} h^{-1} \rm{M}_{\odot}$ units) enclosed within a sphere (SO) and an ellipsoid (EO) of an overdensity of $200\rho_c (z)$, respectively. Sixth and seventh columns correspond to the intermediate-to-major, $(b/c)_{200}$, and minor-to-major, $(a/c)_{200}$, axis ratios of the best-fitting ellipsoid of an overdensity of $200\rho_c (z)$, while the eighth and ninth columns correspond to the intermediate-to-major, $(b/c)_{2500}$, and minor-to-major, $(a/c)_{2500}$, axis ratios of the best-fitting ellipsoid of an overdensity of $2500\rho_c (z)$. For the masses and axis ratios we take into account all particle types (i.e., DM+GAS particles).}
\label{tb:mass}
\begin{tabular}{cccccccccccc}
\hline
\hline
cluster & model & $z$ & $\rm{M_{SO}}$ & $\rm{M_{EO}}$ & $(b/c)_{200}$ & $(a/c)_{200}$ & $(b/c)_{2500}$ & $(a/c)_{2500}$\\
\hline
\hline
clus2 & CDM & 0.250 & $1.88$ & $2.15$ & 0.52 & 0.37 & 0.68 & 0.44\\
clus2 & SIDM & 0.250 & $1.91$ & $2.13$ & 0.54 & 0.42 & 0.81 & 0.71\\
\hline
clus2 & CDM & 0.333 & $2.04$ & $2.21$ & 0.58 & 0.38 & 0.56 & 0.38\\
clus2 & SIDM & 0.333 & $2.05$ & $2.19$ & 0.58 & 0.43 & 0.80 & 0.69\\
\hline
clus2 & CDM & 0.429 & $2.10$ & $2.18$ & 0.64 & 0.44 & 0.56 & 0.35\\
clus2 & SIDM & 0.429 & $2.07$ & $2.16$ & 0.65 & 0.47 & 0.46 & 0.30\\
\hline
clus2 & CDM & 0.538 & $1.82$ & $1.92$ & 0.67 & 0.47 & 0.60 & 0.51\\
clus2 & SIDM & 0.538 & $1.80$ & $1.92$ & 0.66 & 0.48 & 0.76 & 0.72\\
\hline
clus7 & CDM & 0.429 & $1.38$ & $1.49$ & 0.48 & 0.44 & 0.66 & 0.41\\
clus7 & SIDM & 0.429 & $1.36$ & $1.48$ & 0.49 & 0.45 & 0.76 & 0.56\\
\hline
clus9 & CDM & 0.250 & $1.48$ & $1.54$ & 0.70 & 0.47 & 0.59 & 0.40\\
clus9 & SIDM & 0.250 & $1.49$ & $1.54$ & 0.75 & 0.53 & 0.75 & 0.57\\
\hline
clus9 & CDM & 0.333 & $1.36$ & $1.44$ & 0.65 & 0.45 & 0.47 & 0.36\\
clus9 & SIDM & 0.333 & $1.37$ & $1.43$ & 0.73 & 0.53 & 0.62 & 0.53\\
\hline
clus9 & CDM & 0.538 & $1.31$ & $1.34$ & 0.65 & 0.50 & 0.56 & 0.37\\
clus9 & SIDM & 0.538 & $1.32$ & $1.34$ & 0.70 & 0.55 & 0.72 & 0.54\\
\hline
clus11 & CDM & 0.250 & $1.41$ & $1.40$ & 0.60 & 0.55 & 0.26 & 0.20\\
clus11 & SIDM & 0.250 & $1.40$ & $1.39$ & 0.63 & 0.58 & 0.30 & 0.28\\
\hline
clus11 & CDM & 0.300 & $1.31$ & $1.32$ & 0.61 & 0.57 & 0.34 & 0.33\\
clus11 & SIDM & 0.300 & $1.31$ & $1.32$ & 0.68 & 0.64 & 0.45 & 0.43\\
\hline
clus11 & CDM & 0.333 & $1.25$ & $1.28$ & 0.57 & 0.51 & 0.45 & 0.37\\
clus11 & SIDM & 0.333 & $1.24$ & $1.27$ & 0.58 & 0.54 & 0.45 & 0.41\\
\hline
clus30 & CDM & 0.538 & $1.03$ & $1.10$ & 0.59 & 0.45 & 0.41 & 0.38\\
clus30 & SIDM & 0.538 & $1.03$ & $1.10$ & 0.62 & 0.47 & 0.52 & 0.48\\
\hline
clus82 & CDM & 0.429 & $1.03$ & $1.07$ & 0.74 & 0.46 & 0.55 & 0.32\\
clus82 & SIDM & 0.429 & $1.04$ & $1.07$ & 0.74 & 0.48 & 0.56 & 0.39\\

\end{tabular}
\end{center}
\end{table*}

By diagonalizing the mass tensor $M_{\alpha \beta}$ we obtain an initial guess for the shape and orientation of the DM halo. The principal axes of the best-fitting ellipsoid are defined as the square roots of the mass tensor eigenvalues, while their orientations are given by the corresponding eigenvectors. Secondly, we identify the particles located within the ellipsoid defined by the first set of eigenvalues and eigenvectors, and which encloses an ellipsoidal overdensity (EO) equal to $\Delta_c \rho_c (z)$. Then we re-compute the mass tensor for the new distribution of particles to obtain a new set of eigenvalues and eigenvectors that improve the characterization of the halo shape. Finally, we repeat the procedure iteratively until a convergence of a $0.5\%$ in the axis ratios. We denote the minor-to-major axis ratio as $a/c$ and the intermediate-to-major axis ratio as $b/c$, with $a < b < c$. 

\begin{figure}
\centering
 \includegraphics[width=\columnwidth]{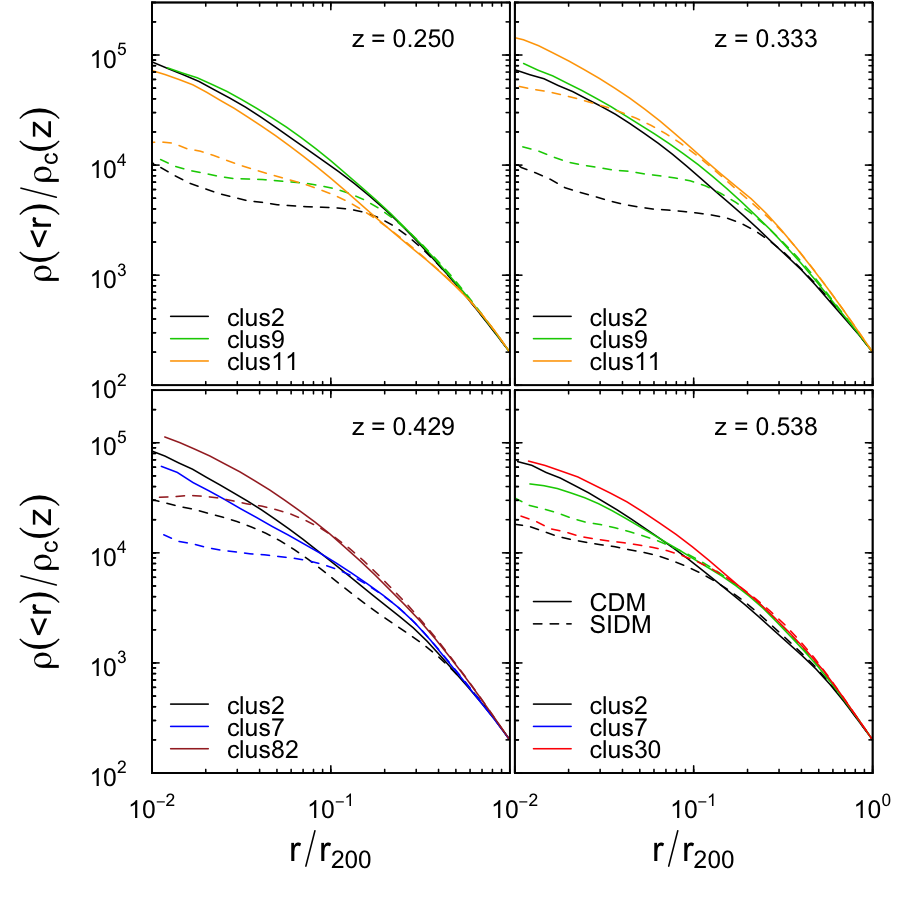}
 \caption{Spherical overdensity profiles of the six massive clusters in our sample. Each panel corresponds to a different redshift, and each color indicates a different cluster. Solid and dashed lines correspond to CDM and SIDM cosmological models, respectively. The total density within a given radius is normalized by the critical density of the Universe at the given redshift, $\rho_c (z)$. Overdensities are computed down to the convergence limit of $\mathrm{r} \sim 20 h^{-1} \mathrm{kpc}$ (roughly $\mathrm{r/r_{200}} \sim 10^{-2}$) and up to a radius equal to $\rm{r}_{200}$.
\label{fig:ovdens}}
\end{figure}

Along with the redshift at which each cluster has been simulated, in table~\ref{tb:mass}, we show the spherical and elliptical virial masses and the axis ratios at the typical overdensity of $\rho(<r) = 200\rho_c (z)$. As found by \citet{Despali2013}, differences between $\rm{M_{SO}}$ and $\rm{M_{EO}}$ are on average about $5\%$ (with $\rm{M_{EO}}$ being systematically larger than $\rm{M_{SO}}$). We found up to a 15$\%$ difference for clus2 at $z = 0.250$ for CDM framework, evidencing the presence of substructures at large radii or an interaction/merger with another DM halo at the given redshift.

Interactions between DM particles are expected to produce more spherical DM halo configurations than collision-less CDM halos, more importantly towards the cluster center where the scattering rate of DM particles is larger \citep{Peter2013,Brinckmann2018}. On average, both the minor-to-major and the intermediate-to-major axis ratios at an overdensity of $200\rho_c (z)$ are systematically larger in the SIDM than in the CDM cosmological model. The median minor-to-major axis ratio of SIDM halos is 7\% higher than in CDM halos. When looking at higher overdensities, i.e., $2500\rho_c (z)$, this ratio increases up to a median value of 1.40, confirming that SIDM simulated cluster-size halos are more rounder than their CDM counterparts. There are some exceptions to this, for instance the SIMD simulation of clus2 at $z = 0.429$ for and overdensity of $2500\rho_c (z)$ is less round, with $(a/c)_{2500} = 0.30$ and $(b/c)_{2500} = 0.46$, than its CDM counterpart, with $(a/c)_{2500} = 0.35$ and $(b/c)_{2500} = 0.56$.

In figure~\ref{fig:ovdens}, we show the spherical overdensity (also cumulative density) profiles of the six massive clusters in our sample at four different redshifts. The cumulative profiles are computed down to the convergence limit \citep[see][for more details]{Power2003}, which roughly corresponds to $\mathrm{r} \sim 20 h^{-1} \mathrm{kpc}$. DM particles self-interactions clearly transform cuspy cores, like those formed in CDM simulations, into flatter cores. The differences are more obvious at lower redshift, when DM particles in massive halos have experienced a larger number of self-interactions, but also DM halos have evolved and underwent a significant number of mergers. The total mass within a radius equal to $\rm{r}_{200}$ remains almost identical in both CDM and SIDM scenarios.

\section{Gravitational lensing properties of DM halos}
\label{sec:lensing}

Interactions between DM particles affect the structure of galaxy clusters, making them more spherical. For large enough DM cross-sections per unit mass (i.e., $\sigma/m \gtrsim 1~\rm{cm}^2/\rm{g}$), these DM interactions may also reduce the number of substructures around them. The same interactions can also transform cuspy cores into flat cores. Since the gravitational lensing is proportional to the projected mass distribution along the line of sight, we expect these differences to arise between the CDM and SIDM halos when comparing the distribution and formation of strong gravitational arcs. Radial arcs can allow us to characterize the size and compactness of the cluster lens cores \citep{Narayan1996}. Moreover, given the more prominent differences between CDM and the SIDM mass profiles at the center of the clusters (as shown in figure~\ref{fig:ovdens}), radial arcs statistics are in principle more sensitive to possible interactions between DM particles.

In order to compare the strong lensing properties for both the CDM and SIDM models, we examine each cluster-size DM halo shown in table~\ref{tb:mass} for 1,000 random orientations along the line of sight to figure out if the halo is super-critical (i.e., if it is able to form critical lines) at each simulated redshift. To do so we use the consolidated ray-tracing code described in \citet{Meneghetti2010} and follow the lensing simulation pipeline assumed in \citet{Meneghetti2017}. All particles belonging to each individual halo are projected along the line-of-sight on the lens plane, while a bundle of light-rays is traced through a regular grid of $512\times512$ covering a region of $250\arcsec \times 250\arcsec$ around the halo center. The deflection angle is computed at each light-ray position after accounting for the contributions from all particles on the lens plane within a box volume of $4.3h^{-1}$Mpc centered at the cluster's center. Then, the deflection field is used to derive lensing quantities, such as the convergence ($\kappa$), the shear ($\gamma$) and the magnification ($\mu$). As described in \cite{Schneider1992}, the lens critical lines are defined as the curves along which the determinant

\begin{equation}
\det A = \mu^{-1} = \mu_t^{-1} \mu_r^{-1} = (1-\kappa-\gamma)(1-\kappa+\gamma) = 0 \;,
\label{eq:jacobian}
\end{equation}
where $A$ is the Jacobian of the lensing potential, $\mu$ is the total magnification, and $\mu_t$ and $\mu_r$ are the tangential and radial magnification, respectively. In particular, the tangential critical line is defined by the condition $\mu_t^{-1} = (1-\kappa-\gamma) = 0$, while the radial critical line occurs when $\mu_r^{-1} = (1-\kappa+\gamma) = 0$. Hereafter, we will use the term \textit{effective} Einstein radius to refer to the size of the tangential critical line which is defined as follows:

\begin{equation}
\theta_E \equiv \frac{1}{D(z_l)}\sqrt{\frac{S}{\pi}}\;,
\label{eq:einsteinr}
\end{equation}
where $S$ is the area enclosed by the tangential critical line and $D(z_l)$ is the angular-diameter distance to the lens plane (see \citealt{Meneghetti2013} and also \citealt{Redlich2012} for more details on the definition of the \textit{effective} Einstein radius).

From these low-resolution maps ($512\times512$ pix$^2$) for each of the 1,000 random projections, we select those showing at least one pixel with a magnification $\mu_r > 1,000$ to ensure that radial critical curves are formed. Given that for real lenses $\kappa$ and $\gamma$ are positive quantities, the condition for forming tangential critical lines is less restrictive than the condition for forming radial critical lines (equation~\ref{eq:jacobian}). Therefore, the formation of radial critical curves implies the formation of tangential critical curves, while the contrary is not true. Consequently, those cases with $\mu_r > 1,000$ are labelled as super-critical (i.e., they show both tangential and radial critical lines).

It is important to note that, for some clusters, none of the projections is super-critical for the SIDM cosmological model, while some of them are super-critical for the CDM cosmology. As we describe below, this effect is more evident for SIDM simulated cluster-size halos as redshift decreases. DM particles interactions throughout cosmic time will gradually dilute the cluster cores. In some massive galaxy clusters, which are formed hierarchically from mergers of DM halos where the number of DM particles interactions is expected to be high, we expect to find cores that are not dense enough to surpass the critical surface mass density for lensing and, therefore, not able to produce either radial or tangential critical lines. In particular, clus2 at $z=0.250$ and $z=0.333$, clus9 at $z = 0.250$ and clus11 at $z = 0.250$ are not super-critical for any of the 1,000 projections along the line of sight for the SIDM cosmological model, while there are some projections for the CDM cosmological model in which they are super-critical. For those cases, the comparison between both cosmological frameworks is not feasible and, therefore, we will not include them in the subsequent analysis. For the remaining clusters and redshifts (i.e., those with super-critical projections in both cosmological models), we re-computed the deflection field and the rest of lensing quantities at higher resolution, $2048\times2048$ pix$^2$ (which translates into an angular resolution of $\approx 0.12$ arcsec in the lens plane), but keeping the rest of parameters constant.

\begin{figure*}
\centering
 \includegraphics[width=2\columnwidth]{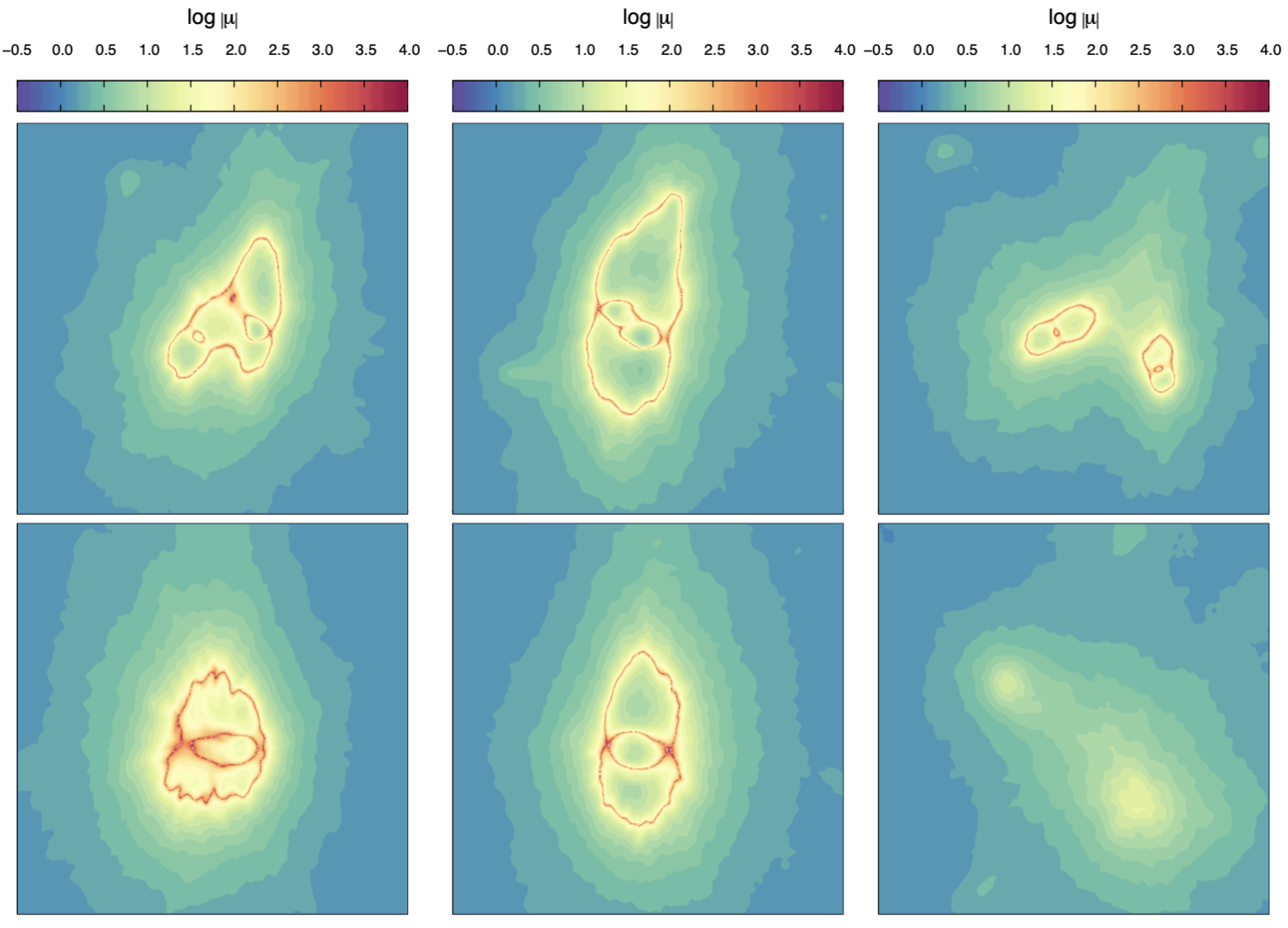}
 \caption{Magnification maps in the lens plane for six randomly selected projections of clus2 at $z=0.429$ (CDM simulations are shown in the top panels, while SIDM simulations are shown in the bottom panels). Color coding denotes the logarithmic of the absolute value of the magnification in the lens plane from $-0.5 < \rm{log}~|\mu| < 4.0$ (and saturated for values of log$~|\mu| > 4$). Both the tangential (outer) and the radial (inner) critical curves are clearly visible for five projections (which are denoted as super-critical), with the exception of the last panel (bottom-right) which shows no critical curves (i.e., it is not super-critical). Moreover, there are some projections showing more than one radial critical curve (e.g., there are three distinguishable radial critical lines in the top-left panel).
\label{fig:clus2_projections}}
\end{figure*}

In figure~\ref{fig:clus2_projections}, we show the magnification maps in the lens plane for six randomly selected projections of clus2 at $z=0.429$: the projections in the top panels correspond to the CDM simulations, while the ones in the bottom panels correspond to the SIDM simulations). There is not a direct correspondence between the top and the bottom panels, although differences between both cosmological models are evident. In particular, for the three projections of the CDM model both tangential and radial critical curves are visible. In the top-left panel, more than one radial critical curve is visible within only one tangential critical curve. This configuration may be explain as two (even three) massive clumps very close in projection, but not enough for the two radial critical lines to merge together (as it is clearly visible in the top-middle panel). On the other hand, when the same two clumps are separate away in projection, it is possible to find two distinct tangential critical lines (with two radial critical lines enclosed by them) due to the two clumps conforming clus2 at $z=0.429$. For the projections of the SIDM model, two of them show both tangential and radial critical lines (i.e., they are super-critical), while the projection in the bottom-right panel is not dense enough to produce tangential or radial critical lines (i.e., it is not super-critical). In particular, for the SIDM simulation of clus2 at $z=0.429$, only 77 projections from the 1,000 random projections produced are super-critical, while the CDM simulation of the same cluster produces 361 super-critical projections out of the 1,000 random projections produced (see table~\ref{tb:arcs}). It is important to remark that these substructure alignments in projection are of course time-dependent and might appear at different times for SIDM than in CDM. Additionally, it is clearly visible how the magnification in areas very close to the radial critical line is higher in SIDM than in CDM projections (wider redder regions). This is due to the expected shallower projected mass profile in the inner regions of SIDM simulations compared to CDM simulations. This flattening in the projected mass profiles also leads to less de-magnified areas within the radial critical line. 

A summary of the number of projections with radial and tangential critical lines for each simulated cluster is shown in table~\ref{tb:arcs}. Overall, for the CDM simulations, the number of projections with tangential and radial critical lines ranges from 148 (clus7 at $z=0.429$) to 771 (clus82 at $z=0.429$), i.e., from a $\sim 15\%$ to a $\sim 80\%$ of the projections, respectively. This fraction drops down to a $\sim 0.1\%$ (clus2 at $z=0.538$) and to a $\sim 40\%$ (clus82 at $z=0.429$) for the SIDM cluster with the lowest and largest number of super-critical projections (both radial and tangential), respectively.


\subsection{Einstein radii statistics}
\label{sec:einstein}

As before mentioned, the size of the Einstein radius is proportional to the total area within the tangential critical line. The tangential critical line is defined by the positions in the lens plane that satisfy $\mu_t^{-1} = 1-\kappa-|\gamma| = 1-\bar{\kappa} = 0$, where $\bar{\kappa}$ is defined as the mean convergence

\begin{equation}
\bar{\kappa} = \frac{\Sigma (<R)}{\Sigma_c (z_l, z_s)}\;,
\label{eq:kappa}
\end{equation}
with $\Sigma (<R)$ is the total projected mass within a given radius (R) normalized by the critical surface density for lensing, $\Sigma_c$, which depends on the lens and source redshifts ($z_l$ and $z_s$, respectively). Therefore, the mean convergence inside a tangential critical curve must equal unity. Consequently, for circularly symmetric lenses, the size of the Einstein radius is proportional to the square-root of the total projected mass within the tangential critical curve.

We compute the two-dimensional maps of $\bar{\kappa}$ and derive the size of the Einstein radius using equation~\ref{eq:einsteinr} for each of the 1,000 random projections along the line of sight produced for each cluster. Hereafter, we fix the source redshift at $z_s = 2.0$. The largest Einstein radius is found to be $\theta_E \approx 63''$ in the CDM simulation of clus11 at $z=0.300$, while its SIDM counterpart shows a slightly smaller value of $\theta_E \approx 58''$. A summary of the Einstein radii statistics is shown in table~\ref{tb:arcs}.

\begin{table*}
\begin{center}
\caption{Summary of gravitational lensing properties. First, second and third columns indicate the cluster, the cosmological model and the clusters' redshift. Fourth column shows the median Einstein radius ($\theta_E$, in arcsec) along with the first and third quantiles. Fifth column corresponds to the largest Einstein radius for each cluster, cosmological model and given redshift. Sixth column indicates the number of projections with tangential critical lines (n$_{\rm{t}}$) from a total of 1,000 projections per cluster and redshift. Seventh column indicates the number of projections with radial critical lines (n$_{\rm{r}}$) from a total of 1,000 projections per cluster and redshift. Eighth column shows the logarithmic of the median length of the radial arcs (log $l$, in arcsec) along with the first and third quantiles. Ninth column corresponds to logarithmic of the length for the largest radial arc (in arcsec). Tenth column shows the logarithmic of the median width of the radial arcs (log $w$, in arcsec) along with the first and third quantiles. Eleventh column corresponds to logarithmic of the width for the widest radial arc (in arcsec). Last column indicates the number of radial arcs identified for each cluster and redshift in the n$_r$ projections with radial critical curves. All the lensing properties have been derived for a source redshift of $z_s = 2.0$.}
\label{tb:arcs}
\begin{tabular}{cccccccccccc}
\hline
\hline
cluster & model & $z$ & $\theta_E ('')$ & max$(\theta_E)$& n$_{\rm{t}}$ & n$_{\rm{r}}$ & log $l ('')$ & max(log $l)$ & log $w ('')$ & max(log $w)$ & n$_{\rm{arcs}}$\\
\hline
\hline
clus11 & CDM & 0.300 & $35.9^{+5.9}_{-12.1}$ & 62.6 & 1000 & 238 & $0.94_{-0.14}^{+0.14}$ & 1.84 & $0.03_{-0.11}^{+0.12}$ & 1.13 & 212219\\[4pt]
clus11 & SIDM & 0.300 & $31.5^{+6.9}_{-12.4}$ & 58.2 & 1000 & 163 & $1.04_{-0.13}^{+0.13}$ & 1.78 & $0.27_{-0.10}^{+0.10}$ & 1.15 & 93961\\[4pt]
\hline
clus11 & CDM & 0.333 & $35.3^{+3.6}_{-3.4}$ & 56.0 & 1000 & 580 & $0.90_{-0.14}^{+0.11}$ & 1.68 & $-0.01_{-0.12}^{+0.13}$ & 1.11 & 343837\\[4pt]
clus11 & SIDM & 0.333 & $20.4^{+13.2}_{-5.5}$ & 50.7 & 1000 & 194 & $1.00_{-0.14}^{+0.13}$ & 1.76 & $0.27_{-0.11}^{+0.11}$ & 1.15 & 67274\\[4pt]
\hline
clus9 & CDM & 0.333 & $20.3^{+3.9}_{-3.2}$ & 41.7 & 966 & 213 & $0.96_{-0.15}^{+0.13}$ & 1.57 & $0.13_{-0.12}^{+0.16}$ & 1.05 & 62213\\[4pt]
clus9 & SIDM & 0.333 & $16.8^{+1.1}_{-2.7}$ & 20.2 & 60 & 37 & $1.03_{-0.17}^{+0.16}$ & 1.72 & $0.77_{-0.15}^{+0.12}$ & 1.19 & 707\\[4pt]
\hline
clus2 & CDM & 0.429 & $21.3^{+14.5}_{-5.5}$ & 52.8 & 1000 & 361 & $0.92_{-0.15}^{+0.13}$ & 1.69 & $0.04_{-0.14}^{+0.16}$ & 1.05 & 135897\\[4pt]
clus2 & SIDM & 0.429 & $33.2^{+2.2}_{-3.8}$ & 49.1 & 297 & 77 & $1.04_{-0.17}^{+0.13}$ & 1.59 & $0.21_{-0.12}^{+0.16}$ & 1.19 & 19173\\[4pt]
\hline
clus7 & CDM & 0.429 & $19.5^{+7.9}_{-4.9}$ & 49.4 & 802 & 148 & $0.91_{-0.14}^{+0.11}$ & 1.63 & $0.19_{-0.14}^{+0.15}$ & 1.15 & 37496\\[4pt]
clus7 & SIDM & 0.429 & $18.0^{+2.6}_{-5.5}$ & 41.4 & 335 & 40 & $1.09_{-0.17}^{+0.16}$ & 1.71 & $0.55_{-0.19}^{+0.19}$ & 1.20 & 3291\\[4pt]
\hline
clus82 & CDM & 0.429 & $33.2^{+4.7}_{-5.5}$ & 45.0 & 1000 & 771 & $0.86_{-0.14}^{+0.11}$ & 1.71 & $-0.07_{-0.13}^{+0.15}$ & 1.11 & 437454\\[4pt]
clus82 & SIDM & 0.429 & $30.4^{+4.6}_{-4.7}$ & 43.6 & 1000 & 449 & $0.97_{-0.16}^{+0.13}$ & 1.62 & $0.13_{-0.12}^{+0.14}$ & 1.20 & 206500\\[4pt]
\hline
clus2 & CDM & 0.538 & $18.8^{+4.6}_{-2.8}$ & 32.3 & 733 & 411 & $0.86_{-0.15}^{+0.13}$ & 1.61 & $0.14_{-0.14}^{+0.14}$ & 1.01 & 69457\\[4pt]
clus2 & SIDM & 0.538 & $18.4^{+3.6}_{-3.4}$ & 29.5 & 160 & 14 & $1.01_{-0.21}^{+0.19}$ & 1.46 & $0.44_{-0.17}^{+0.13}$ & 1.11 & 619\\[4pt]
\hline
clus9 & CDM & 0.538 & $27.2^{+2.1}_{-4.1}$ & 40.5 & 943 & 497 & $0.91_{-0.15}^{+0.14}$ & 1.65 & $-0.02_{-0.12}^{+0.15}$ & 1.00 & 166090\\[4pt]
clus9 & SIDM & 0.538 & $23.0^{+2.1}_{-4.8}$ & 33.7 & 869 & 301 & $0.99_{-0.16}^{+0.14}$ & 1.63 & $0.29_{-0.13}^{+0.16}$ & 1.16 & 42223\\[4pt]
\hline
clus30 & CDM & 0.538 & $21.1^{+3.7}_{-3.5}$ & 37.2 & 1000 & 272 & $0.85_{-0.14}^{+0.12}$ & 1.67 & $0.08_{-0.12}^{+0.13}$ & 0.98 & 100856\\[4pt]
clus30 & SIDM & 0.538 & $17.3^{+5.6}_{-4.7}$ & 33.9 & 602 & 427 & $0.98_{-0.16}^{+0.13}$ & 1.60 & $0.50_{-0.14}^{+0.14}$ & 1.16 & 19179\\

\end{tabular}
\end{center}
\end{table*}

\subsection{Radial arcs}
\label{sec:radial}

\begin{figure*}
\centering
 \includegraphics[width=0.6\columnwidth]{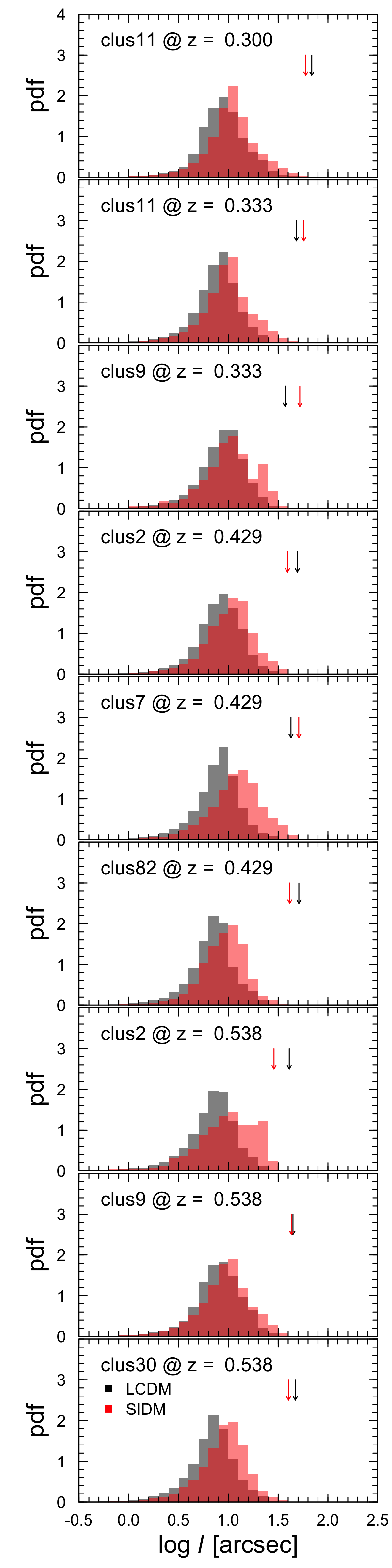}
 \includegraphics[width=0.6\columnwidth]{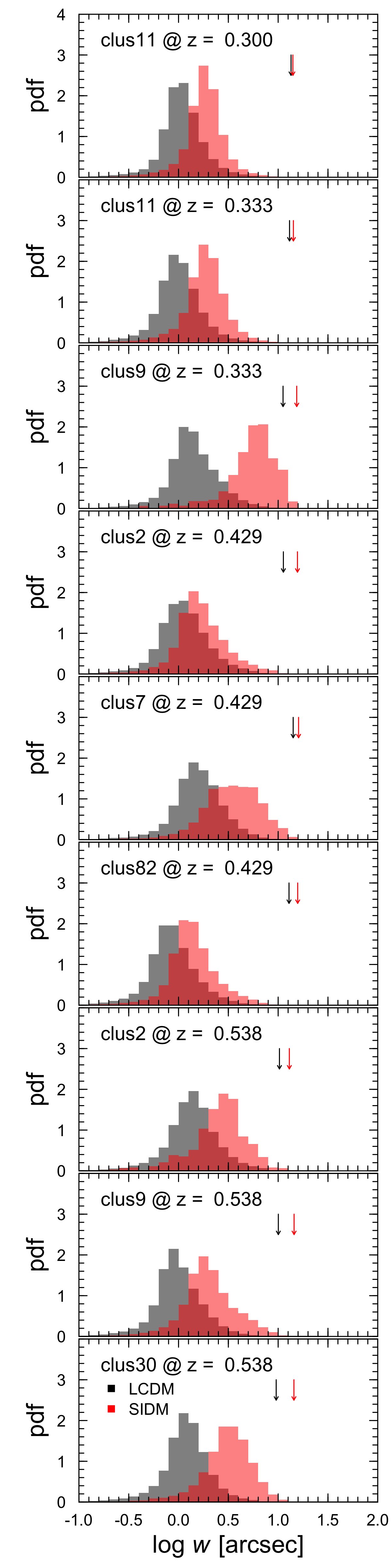}
 \includegraphics[width=0.6\columnwidth]{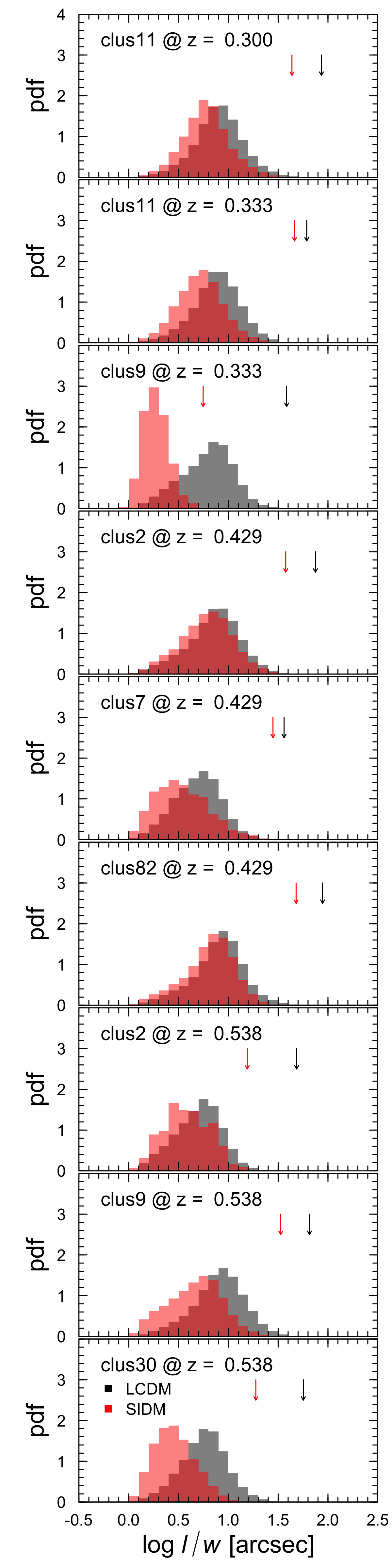}
 \caption{Probability distribution functions of lengths ($l$, left-hand panels), widths ($w$, middle panels) and length-to-width ratios ($l/w$, right-hand panels) in log-scale for the selected radial arcs. Each panel corresponds to one cluster and redshift (indicated by the label in the top-left corner of each panel). Grey and red histograms show the results for CDM and SIDM cosmological models, respectively, while the black and red arrows correspond to the maximum value found in each case.
\label{fig:pdfs}}
\end{figure*}

The formation and location of radial arcs by galaxy clusters depends both on the slope of the projected mass profile and on the central density of the lens \citep{Narayan1996,Meneghetti2013}. Overall, the conclusion holds that the presence of radial arcs indicate that clusters have dense cores with fairly flat density profiles \citep[see e.g.][]{Kormann1994}. Moreover, the steeper the density profile, the closer to the centre the radial arcs tend to be located \citep{Williams1999}. As shown in figure~\ref{fig:ovdens}, DM particles self-interactions favor the formation of flat cores in massive clusters and, therefore, the formation of large radial arcs if they are dense enough (i.e., $\mu_r^{-1} = 1-\kappa+|\gamma| \approx 0$). Background objects, such as high-$z$ galaxies, located close to the caustics (i.e., the analogs to the critical lines in the source plane) will appear strongly distorted (and/or multiply imaged) in the lens plane. The caustics can be computed by mapping the critical lines onto the source plane using the lens equation.

In this subsection, we present an analysis on the distribution of large radial arcs produced by the clusters and redshifts with at least one super-critical projection in both cosmological models. We follow a procedure similar to the one presented by \citet{Meneghetti2001} to produce lensed images of background sources. The sources are initially distributed on a regular grid in the source plane. Subsequently, the spatial density of the sources iteratively increases near the caustics in order to obtain a larger number of them placed on regions where the probability of being strongly distorted is higher. Sources are assumed to be elliptical, with axis ratios uniformly distributed in the interval $(0.5,1)$, and with an area equal to a circle of 1 arcsec diameter. The resolution of $0.12$ arcsec of the gravitational lensing maps is large enough to properly resolve the lensed images of the background sources. Then, we classify as strong radial arcs those lensed images of background sources containing at least one pixel for which $\mu_r > 5$ and $\mu_r / \mu_t > 4$. These choices were made to avoid miss-classifications of radial arcs, such as arcs that can be considered both radial and tangential at the same time as the radial and critical curves overlap.

\subsubsection{Lengths, widths and length-to-width ratios}
\label{sec:radial_shapes}

We characterize the lensed images classified as radial arcs in terms of their lengths and widths following the procedure introduced by \citet{Bartelmann1994} and described in \citet{Meneghetti2013}. The lengths are defined as the maximum length of the circular segment passing through the lensed image, while the widths are found by fitting the image with several geometric forms (such as ellipses, circles, rectangles and rings).

Given that SIDM cluster-size halos generally have shallower profiles than CDM ones towards their inner parts, we expect them to produce more elongated radial arcs. However, as we showed in section~\ref{sec:shapes}, CDM cluster-size halos are more triaxial at their inner parts, compensating for the shallower mass profiles of SIDM halos and explaining why the lengths of radial arcs are comparable in both cosmological models. In figure~\ref{fig:pdfs}, we show the distributions in lengths ($l$), widths ($w$) and length-to-width ratios ($l/w$) for the radial arcs in each cluster and redshift. Each panel includes all the radial arcs identified in each of the projections along the line of sight having radial critical lines. The total number of radial arcs for each cluster, redshift and cosmological model along with a summary with their statistics of lengths and widths are shown in table~\ref{tb:arcs}. The number of projections with radial critical lines and the total number of radial arcs are systematically larger for the CDM than for the SIDM cosmological models. Nevertheless, the overall trend is that the distribution of lengths for both cosmological models is equivalent within errors for the six cluster-size halos and each corresponding redshift here presented. The median values for the lengths of radial arcs are slightly larger for the SIDM than for the CDM model. The largest radial arc, $l\approx 70$ arcsec, is produced by one of the projections of clus11 at $z=0.300$ simulated with CDM cosmology. This extremely elongated radial arcs is the result of the merger of five multiple images into a single one due to the superposition of several radial caustics along the source. In particular, for the projection of clus11 producing this particular radial arc it is possible to identify three different radial critical lines which are similar (in the lens plane) to the top-left panel in figure~\ref{fig:clus2_projections}. Although radial arcs are more difficult to be characterized than tangential arcs since they appear close to the cluster centers and, therefore, their light is usually screened by the BCG light, there have already been identified large radial arcs in observed clusters, such as the radial arc found by \citet{Caminha2017} in MACS J1206 (system 4b with an approximately length of 10 arcsec).

Contrarily to the findings of \citet{Meneghetti2001} for only one simulated cluster-sized halo, which constrained $\sigma/m < 0.1$ cm$^2/$g for it to produce extreme strong lensing arcs, each of the clusters here presented is able to produce giant radial arcs (also large Einstein radii) assuming $\sigma/m = 1$ cm$^2/$g within the SIDM cosmological model. One of the reasons of the simulated halo in \citet{Meneghetti2001} for not being dense enough to produce strong lensing features may be the redshift at which the halo has been analyzed, $z=0.278$. As mentioned above, we find that some of the clusters at lower redshifts (i.e., $z=0.250$ for clus2, clus9 and clus11, and $z=0.333$ for clus2) are not massive and dense enough to produce tangential and radial critical curves. Depending on the mass accretion and merger histories of each cluster-size halo, as they evolve with redshift, DM self-interactions lead to a dilution of the DM cores and, consequently, prevents them from being strong gravitational lenses (i.e., super-critical). However, we find that both the CDM and the SIDM cosmological models lead to the formation of extremely large Einstein radii and giant radial arcs for all the halos analyzed at higher redshifts ($z = 0.429$ and $z=0.538$) and for clus11 at $z = 0.300$ and $z = 0.333$. When comparing at the distributions of the width of radial arcs in the two cosmological models, the situation is slightly different. The radial arcs produced by the same cluster-size halo are on average wider for the SIDM, with median values in the range log $w = (-0.13, 0.77)$, than for the CDM cosmological model, with median values in the range log $w = (-0.07, 0.19)$. Since the radial arcs formed by the SIDM clusters tend to be wider while approximately of the same length, we consequently find that length-to-width ratios are systematically smaller in SIDM than in CDM simulations. In particular, clus9 at $z = 0.333$ in SIDM forms radial arcs with similar lengths but significantly wider than in CDM and, therefore, with considerably smaller values for the length-to-width ratios.

\subsubsection{Convergence, shear and radial magnification}
\label{sec:mu_radial}

\begin{figure}
\centering
 \includegraphics[width=\columnwidth]{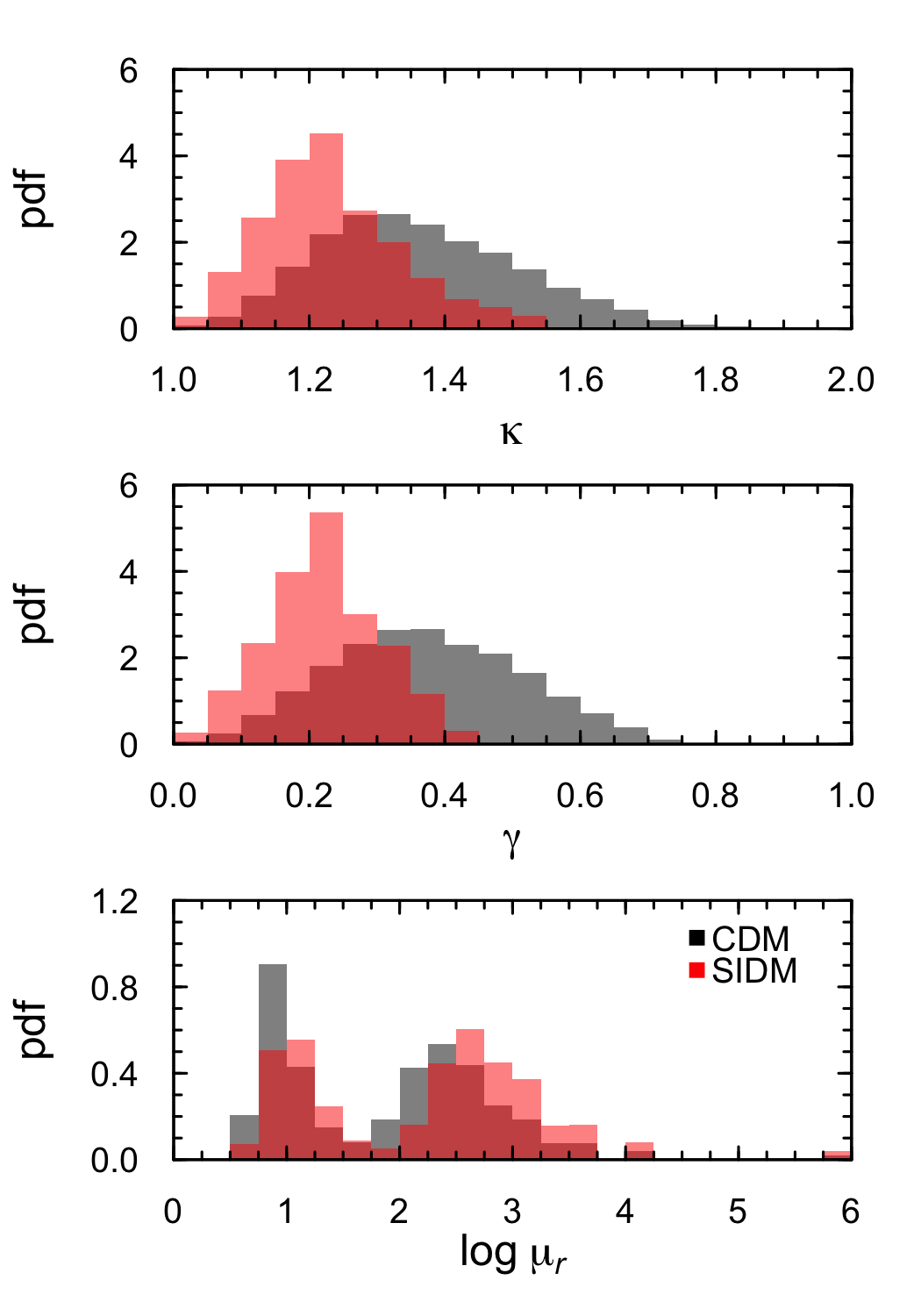}
 \caption{Probability distribution functions of convergence ($\kappa$, top panel), shear ($\gamma$, middle panel) and the logarithmic of the radial magnification ($\rm{log}~\mu_r$, bottom panel) for the selected radial arcs in all clusters and redshifts. Grey and red histograms show the results for CDM and SIDM cosmological models, respectively.
\label{fig:mu_all}}
\end{figure}

In order to highlight the differences between the radial arcs formed by the different clusters in the CDM and SIDM cosmological models, we also derived the convergence ($\kappa$), the shear ($\gamma$) and, as a combination of both, the radial magnification ($\mu_r^{-1} = 1-\kappa+|\gamma|$) for each radial arc. The values of $\kappa$ and $\gamma$ are computed in the central pixel of the lensed images in the lens plane for each of the detected radial arcs.

In figure~\ref{fig:mu_all}, we show the distributions in $\kappa$, $\gamma$ and $\rm{log}~\mu_r$ for all the radial arcs identified in the six clusters and the different redshifts (the total number of radial arcs is equal to the sum of the column n$_{\rm{arcs}}$ in table~\ref{tb:arcs} for each cosmological model). The radial arcs formed by clusters simulated with SIDM cosmological model show values of $\kappa$ systematically smaller than those produced by the same clusters simulated within the CDM framework. The median values of the convergence (along with the 1st and 3rd quartiles) for the radial arcs are $\bar{\kappa}_{\Lambda\rm{CDM}} \simeq 1.35^{+0.12}_{-0.09}$ and $\bar{\kappa}_{\rm{SIDM}} \simeq 1.22^{+0.07}_{-0.06}$ for CDM and SIDM cosmological models, respectively. The maximum values of the convergence for CDM and SIDM cosmological models are approximately 1.96 and 1.59, respectively. When looking at the values of the shear, radial arcs identified in clusters simulated with SIDM cosmological model tend to be formed in regions with smaller values of $\gamma$. More precisely, the median values of the shear (along with the 1st and 3rd quartiles) for the radial arcs are $\bar{\gamma}_{\Lambda\rm{CDM}} \simeq 0.37^{+0.11}_{-0.10}$ and $\bar{\gamma}_{\rm{SIDM}} \simeq 0.22^{+0.06}_{-0.05}$ for CDM and SIDM cosmological models, respectively. Finally, the pdfs of the radial magnification ($\mu_r$) are consistent in both cosmological models, with a preference for slightly higher values ($\rm{log}~\mu_r > 2.5$) in the SIDM framework.

\section{Conclusions}
\label{sec:discussion}

We compare the statistics and morphology of extremely large radial arcs produced by a set of six simulated cluster-size DM halos. The simulated galaxy clusters of study are selected for being the most efficient gravitational lenses found in two datasets of re-simulated galaxy clusters with slightly different cosmological parameters but the same simulated cubic box volume of $1h^{-1}$Gpc side. The six selected galaxy clusters are then simulated using the N-body/SPH framework GIZMO assuming a CDM and a SIDM cosmological model with a velocity independent cross-section for the DM particles of $\sigma / m = 1$ cm$^2/$g. Finally, we study the gravitational lensing properties for 1,000 random orientations along the line sight of each simulated cluster-size halo using a ray-tracing pipeline by selecting those producing both radial and tangential critical lines for a given source redshift ($z_s = 2.0$). To produce the lensing images of background sources by these simulated halos we populate the source plane behind them with elliptical sources. We then select those lensed images that are classify as strong radial arcs according to their radial and tangential magnifications, and derive the probability distributions of their lengths, widths and radial magnifications.

By looking at the overall properties of the cluster-size halos (see table~\ref{tb:mass}), we found that the axis ratios measured at an overdensity of $200\rho_c (z)$ are systematically (but only slightly) larger in SIDM than in CDM simulations of the same cluster-size halos. If we refer to the same axis ratios but measured at an inner radius (e.g., at an overdensity of $2500\rho_c (z)$) the differences are significantly larger, with a value of 1.4 for the median of the ratio of the minor-to-major axis ratios of SIDM and CDM, confirming that SIDM simulated cluster-size halos are on average rounder than their CDM counterparts at an overdensity of $2500\rho_c (z)$. As expected, DM particles self-interactions clearly transform cuspy cores (like those formed in CDM simulations) into flat cores (as found in SIDM simulations), as is clearly seen in figure~\ref{fig:ovdens}.

The gravitational lensing properties are derived by examining 1,000 random projections for each cluster, redshift and cosmological model, and assuming a source redshift $z_s = 2.0$. We define as super-critical the projections for which both tangential and radial critical lines are present. We found no super-critical projections for any of the SIDM simulations at $z=0.250$, indicating that some massive galaxy clusters for which the number of DM particles interactions are expected to be high are not able to produce either radial or tangential critical lines. Contrarily, even for the lower redshift ($z=0.250$), at least one of the projections of the CDM simulations is super-critical. To illustrate this effect we show the magnification maps for six random projections of clus2 at $z=0.429$ in figure~\ref{fig:clus2_projections}, where (depending on the projection) one, two and even three radial critical curves could be detected. Additionally, the values of the magnification along and within the radial critical lines for the SIDM simulation are larger than in the CDM simulations.

The clusters and redshifts for which super-critical projections in both cosmological models exist, are then simulated with higher resolution ($\approx 0.12$ arcsec) to study in detail the Einstein radii and the radial arc statistics. The overall statistics of the gravitational lensing properties are shown in table~\ref{tb:arcs}. The main conclusions are summarized as follows:

\begin{itemize}

    \item The fraction of projections that result in critical lines is systematically larger in the CDM than in the SIDM cosmological model. In particular, the fraction projections for the CDM simulations that produce both both tangential and radial critical lines ranges between $15\%$ and $80\%$ depending on the cluster and the redshift, while this fraction varies between $0.1\%$ and $40\%$ for the SIDM simulations.

    \item Although some of the clusters at lower redshifts (i.e., $z=0.250$ for clus2, clus9 and clus11, and $z=0.333$ for clus2) are not massive and dense enough to produce tangential and radial critical curves, we find that both the CDM and the SIDM cosmological models lead to the formation of extremely large Einstein radii and giant radial arcs for all the halos analyzed at higher redshifts (i.e., $z = 0.429$ and $z=0.538$) and for clus11 at $z=0.300$ and $z=0.333$. In particular, the largest Einstein radius is found to be $\theta_E \approx 63''$ in the CDM simulation of clus11 at $z=0.300$, while its SIDM counterpart shows a slightly smaller value of $\theta_E \approx 58''$. The largest radial arc, $l\approx 70$ arcsec, is produced by one of the projections of clus11 at $z=0.300$ simulated with CDM cosmology.
    
    \item The distribution of lengths for both cosmological models is equivalent within errors for the six cluster-size halos and each corresponding redshift here presented, with median values for the lengths of radial arcs slightly larger for the SIDM than for the CDM model.
 
    \item Radial arcs produced by the same cluster-size halo are on average wider for the SIDM, with median values in the range log $w = (-0.13, 0.77)$, than for the CDM cosmological model, with median values in the range log $w = (-0.07, 0.19)$.
    
    \item Radial arcs formed by clusters simulated with SIDM cosmological model show values of $\kappa$ systematically smaller than those produced by the same clusters simulated within the CDM framework. The median values of the convergence (along with the 1st and 3rd quartiles) for the radial arcs are $\bar{\kappa}_{\Lambda\rm{CDM}} \simeq 1.35^{+0.12}_{-0.09}$ and $\bar{\kappa}_{\rm{SIDM}} \simeq 1.22^{+0.07}_{-0.06}$ for CDM and SIDM cosmological models, respectively. The maximum values of the convergence for CDM and SIDM cosmological models are approximately 1.96 and 1.59, respectively.
    
    \item Radial arcs identified in clusters simulated with SIDM cosmological model tend to be formed in regions with smaller values of $\gamma$. More precisely, the median values of the shear (along with the 1st and 3rd quartiles) for the radial arcs are $\bar{\gamma}_{\Lambda\rm{CDM}} \simeq 0.37^{+0.11}_{-0.10}$ and $\bar{\gamma}_{\rm{SIDM}} \simeq 0.22^{+0.06}_{-0.05}$ for CDM and SIDM cosmological models, respectively.
    
    \item Finally, the pdfs of the radial magnification ($\mu_r$) are consistent in both cosmological models, with a preference for slightly higher values ($\rm{log}~\mu_r > 2.5$) in the SIDM cosmological model.
    
\end{itemize}

In cluster-scale lenses, parametric lens models are typically inferred using the positional information of the lensed images and adopting CDM-inspired parametric models. These models can later be used to infer the predicted shape (i.e., length and width) of the radial arcs. Deviations between predicted and observed shapes of radial arcs can point to possible tensions with the adopted CDM model (usually a NFW profile or variations of an spherical isothermal model). In these cases, alternative models such as SIDM should be considered, and the widening of radial arcs in our SIDM simulations can be exploited to constrain the cross-section in this type of models. Although difficult to model, baryonic effects can modify the central potential affecting the statistics and shape of the innermost lensed images. However, the impact of the baryon component can be easily incorporated into lens models by adopting a mass component that traces the observed luminous matter.

In this context, this work demonstrates that simulated cluster lenses within the SIDM framework with a velocity independent cross-section for the DM particles of $\sigma / m = 1$ cm$^2/$g are capable to form extended radial arcs with comparable (or ever higher) radial magnifications with respect to the CDM cosmological model. Therefore, we may conclude that it is not possible to rule out a cross-section for the DM particles of $\sigma / m \lesssim 1$ cm$^2/$g based on the formation of radial arcs by simulated galaxy clusters, as previously done by \citet{Meneghetti2001}. Moreover, we show how the shape of radial arcs contains valuable information about the densest central region of galaxy clusters, where possible interactions between DM particles are more likely to take place. Detailed analyses of observed radial arcs in present and future high-resolution data can be used to set further limits on the cross-section of SIDM cosmological models.

\section*{Acknowledgments}
The authors would like to thank the ``Red Espa\~nola de Supercomputaci\'on" for granting us computing time at the MareNostrum Supercomputer of the BSC-CNS where some of the cluster simulations presented in this work have been performed. GY acknowledges financial support by the MINECO/FEDER under project grant AYA2015-63810-P and MICIU/FEDER under project grant PGC2018-094975-C21. WC acknowledges the supported by the European Research Council under grant number 670193. MM acknowledges support from PRIN-MIUR ``Cosmology and Fundamental Physics: illuminating the dark universe with Euclid", and from ASI through contract Euclid Phase D 1.05.04.37.01. 

\section*{Data availability}
The simulated cluster-scale DM halos presented in this article will
be shared on request to the corresponding author. A public version of the GIZMO code (written and maintained by Philip F. Hopkins) used to perform the cluster simulations is available at: \url{https://bitbucket.org/phopkins/gizmo-public/src/master/}. The ray-shooting pipeline applied on the cluster simulations to derive their gravitational lensing properties was provided by M. Meneghetti and will be shared on request to the corresponding author with permission of M. Meneghetti.
\bibliographystyle{mnras}
\bibliography{radial_sidm}


\bsp	
\label{lastpage}
\end{document}